\documentclass[prl,twocolumn,floatfix]{revtex4}
\usepackage{amsmath}
\usepackage{amsfonts}
\usepackage{amssymb}
\usepackage{graphicx}

\begin{document}

\title{Surface Plasmon Assisted Kondo Resonances on a Metallic Nanowire}
\author{Ren-Shou Huang}
\affiliation{Research Center for Applied Sciences, Academia Sinica, Taipei, Taiwan 11529, R.O.C.}
\author{Yia-Chung Chang}
\affiliation{Research Center for Applied Sciences, Academia Sinica, Taipei, Taiwan 11529, R.O.C.}
\date{\today}
\bibliographystyle{apsrev}

\begin{abstract}
In this letter we  propose an experiment to measure the Kondo effect
for magnetic atoms adsorbed on the surface of a metallic nanowire.
In addition to the traditional $sp$-$d$ hybridization, by
introducing the strong electromagnetic field of the localized
surface plasmon on the nanowire, we show that it is possible to
observe additional $sp$-$d$ electron transfer processes assisted by
surface plasmons. Due to the good surface-to-volume ratio of the
nanowire, the Kondo resonances here would be revealed as multiple
anti-resonances in the differential conductance versus bias voltage
curve.
\end{abstract}

\maketitle

The Anderson model was introduced initially to explain the localized
magnetic moments in metals\cite{Anderson61}. The Coulomb repulsion
on the impurities for localized orbitals provides a great variety of
interesting physics. One of the important features is the appearance
of the Kondo resonance near the Fermi surface. It leads to many
unusual effects particularly in strongly correlated electronic
systems. More recently the scanning tunneling microscopy measurement
of a single magnetic atom on a metallic surface\cite{Madhavan98} was
a direct observation of Kondo resonance.

Meanwhile, \emph{surface plasmon resonance} (SPR), which has been
known for a long time to give rise to various colors in fine noble
metal particles, is an effect that can be understood in classical
electromagnetic theory\cite{Maxwell-Garnett04,Mie08}. Its amplitude
of EM field is strongly enhanced due to the fact that it is highly
localized near metal surfaces. Also depending on the geometry,
surface plasmon can be non-radiative in some cases and therefore
travel a long distance.\cite{Sarid81} These special characteristics
have allowed SPR to find its ways to many applications in different
areas, such as surface enhanced Raman scattering\cite{Moskovits85}.

\begin{figure}
\centering
\includegraphics[scale=0.4]{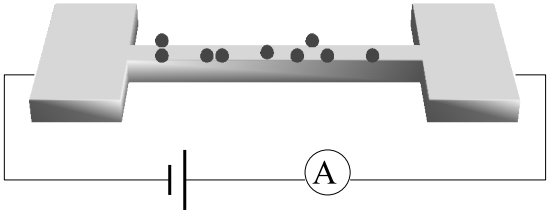}
\caption{Illustration of the experimental setup. T
he suspended metal nanowire is connected to an electrode at each end for the measurement of the conductance.
The Kondo atoms such as Co are chemically adsorbed on the wire.
The surface plasmon can be excited by shining laser on one of the electrodes\cite{Weeber03}.}\label{fig:nanowire}
\end{figure}
In this letter we propose an experiment as illustrated in
Fig.~\ref{fig:nanowire}. The use of metallic nanowire serves two
purposes here. One is to maximize the scattering of the electrons by
the surface of the wire, so that the electrons going through the
wire are more sensitive to the Kondo atoms on the surface. The other
is for introducing surface plasmons. Since surface plasmons only
exist near the surface, the localization of the EM field caused by
the confinement in the transverse directions can ensure the large EM
enhancement. The strong EM field couples to electrons through the
$\mathbf{j}\cdot\mathbf{A}$ interaction. The excitation of the
surface plasmon can be carried out in one of the electrodes. It has
been shown that surface plasmon can travel coherently up to 10
$\mu$m in a single-crystalline silver nanowire\cite{Ditlbacher05}.

The Hamiltonian considered is
\begin{eqnarray}
H & = & H_C+H_T;\nonumber\\
H_C & = & \sum_{\mathbf{k}\sigma}\epsilon_\mathbf{k}c_{\mathbf{k}\sigma}^\dag c_{\mathbf{k}\sigma}+\epsilon_d\sum_{i=1}^n\sum_{\nu=1}^{N_d}d_{i\nu}^\dag d_{i\nu}+U\sum_{i\nu\nu'}n_{id\nu}n_{id\nu'}\nonumber\\
& & +\sum_\mathbf{q}\hbar\omega_\mathbf{q}a_\mathbf{q}^\dag a_\mathbf{q}+\sum_{\mathbf{k}\sigma i\nu\mathbf{q}}V_{\mathbf{k} i\mathbf{q}}c_{\mathbf{k}\sigma}d_{i\nu}^\dag(a_\mathbf{q}+a_\mathbf{q}^\dag)+\textrm{hc},\nonumber\\
H_T & = & \sum_{j=L,R}\sum_{\mathbf{k}\sigma}(\epsilon_\mathbf{k}+\mu_j)b_{j\mathbf{k}\sigma}^\dag b_{j\mathbf{k}\sigma}\nonumber\\
& & \qquad+\sum_{j=L,R}\sum_{\mathbf{kk'}\sigma}t_{j\mathbf{kk'}}c_{\mathbf{k}\sigma}b_{j\mathbf{k'}\sigma}^\dag+\textrm{hc}.
\end{eqnarray}
$H_C$ describes the Anderson model for a nanowire with $n$
impurities coupled to assisting photons. $c_{\mathbf{k}\sigma}$ and
$d_{i\nu}$ are the annihilation operators for the electrons on the
wire and on the $i$th atom. $a_\mathbf{q}$ is the annihilation
operator for the surface plasmon. The indices $\mathbf{k}$ and
$\mathbf{q}$ are assumed continuous in the longitudinal direction
and discrete in the transverse directions of the wire. The inelastic
electron-surface plasmon scattering term comes from the
$\mathbf{j}\cdot\mathbf{A}$ interaction, where $\mathbf{j}$ is the
tunneling current between the wire and the atom. The same
$\mathbf{j}\cdot\mathbf{A}$ interaction for the current inside the
wire has been integrated into the self-energy of the EM field which
gives rise to the negative dielectric constant in metal and forms
the surface plasmon\cite{Pines}. The tunneling term in the original
Anderson model for the $sp$-$d$ hybridization is included in the
$\mathbf{q}=0$ term, where $\omega_{\mathbf{q}=0}=0$. $H_T$
describes the coupling of the nanowire to the identical electrodes
at both ends. $b_{L\mathbf{k}\sigma}$ and $b_{R\mathbf{k}\sigma}$
denote the annihilation operators for the electrons on the left and
the right lead, respectively. $\mu_L$ and $\mu_R$ are the chemical
potentials of the fermi seas on the leads.

The surface plasmon field on the nanowire in general has two kinds
of modes, leaky modes and bound modes.\cite{Zia05}. In both cases,
the condition for surface plasmon resonance derived from classical
Mie scattering theory is \(
\textrm{Re}[\varepsilon_m(\omega=\omega_\mathrm{SPR})]=-\varepsilon_0.
\) $\varepsilon_m$ is the dielectric constant of the metal, and
$\varepsilon_0$ is that of the surrounding medium which is real and
positive. The mechanisms for the energy loss in the two kinds of
modes are different. The leaky modes lose energy mostly due to
irradiation. For the bound modes it is mostly caused by the internal
dissipation of the metal characterized by
$\textrm{Im}[\varepsilon_m]$, which classically can be attributed to
the Drude damping. However, when the size of the nanostructure is
considerably smaller than the electron mean free path, the
dissipation of both bound modes and leaky modes is dominated by
$\textrm{Im}[\varepsilon_m]$ as a result of the electrons'
scattering with the surface\cite{Kawabata66}. In either case, the
surface plasmon frequency dependence of the field amplitude, or
equivalently the electron-surface plasmon coupling constant, can be
approximated as
\begin{equation}
V_{\mathbf{k}i\mathbf{q}}\simeq e^{i\mathbf{k}\cdot\mathbf{r}_i}V_\mathbf{q}
=\left(V\frac{i\kappa/2}{\omega_\mathbf{q}-\omega_\mathrm{SPR}+i\kappa/2}
+V_0\delta_{\mathbf{q},0}\right)e^{i\mathbf{k}\cdot\mathbf{r}_i},
\end{equation}
with $\kappa$ the full width of the resonance. The electron momentum
dependence of the coupling constant enters with different locations
of the Kondo atoms.  $V_0$ describes the original $sp$-$d$
hybridization between magnetic adatom and the metallic nanowire and
$V$ describes the coupling of the $sp$-$d$ current with the surface
plasmon field. For an isolated gold nanowire of diameters $\sim30$
nm and micrometers length, the surface plasmon has a resonance at
$\sim525$ nm free space wavelength and a full width of $\sim50$
nm\cite{Mock02}. The intensity of the local electric field is
estimated to exceed the incident field by $10^3$\cite{Podolskiy03}.

The DC current on the nanowire is given by the Landauer formula in an interacting electron region\cite{Meir92}
\begin{equation}
J=-\frac{2e}{h}\int d\epsilon[f_L(\epsilon)-f_R(\epsilon)]\textrm{Im}[\textrm{tr}\{\Gamma G^r\}],
\end{equation}
where $f_L$ and $f_R$ is the Fermi distribution functions on the left and the right leads, $G^r\equiv G_\mathbf{k\sigma k'\sigma'}^r$ is the full retarded Green's function for the electrons on the wire, and $\Gamma\equiv\Gamma^L\Gamma^R/(\Gamma^L+\Gamma^R)$. $\Gamma^L$ is the electron tunneling rate to the left lead and is defined as $\Gamma_\mathbf{k\sigma k'\sigma}^L=2\pi\sum_\mathbf{k''}\rho_\mathbf{k''}^Lt_\mathbf{kk''}t_\mathbf{k''k'}^*$, so is $\Gamma^R$. $\rho_\mathbf{k''}^L$ is the density of states in the left lead. If we assume that $\Gamma^L$ and $\Gamma^R$ are constants, then the expression is simplied to
\begin{equation}
J=-\frac{2e\Gamma}{h}\int d\epsilon[f_L(\epsilon)-f_R(\epsilon)]\sum_{\mathbf{k}\sigma\mathbf{k'}\sigma'}\textrm{Im}\left[G_{\mathbf{k}\sigma\mathbf{k'}\sigma'}^r(\epsilon)\right].\label{eq:J}
\end{equation}

Here we expand the retarded Green's function to second order
\begin{eqnarray}
&&\sum_{\mathbf{k}\sigma\mathbf{k'}\sigma'}G_{\mathbf{k}\sigma\mathbf{k'}\sigma'}^r(\epsilon)
\simeq
\sum_{\mathbf{k}\sigma}G_{\mathbf{k}\sigma\mathbf{k}\sigma}^0  \nonumber \\
&&+\sum_{i=1}^n\sum_{\mathbf{k}\sigma\mathbf{k'}\sigma'}G_{\mathbf{k}\sigma\mathbf{k}\sigma}^0
e^{i\mathbf{k}\cdot\mathbf{r}_i}F(\epsilon)e^{-i\mathbf{k'}\cdot\mathbf{r}_i}
G_{\mathbf{k'}\sigma'\mathbf{k'}\sigma'}^0;\label{eq:G}\\
&& G_{\mathbf{k}\sigma\mathbf{k}\sigma}^0\equiv\frac{1}{\epsilon-\epsilon_\mathbf{k}+i(\Gamma^L+\Gamma^R)/2}, \nonumber \\
&&\rho_0=-\frac{1}{\pi}\sum_{\mathbf{k}\sigma}\mathrm{Im}[G_{\mathbf{k}\sigma\mathbf{k}\sigma}^0], \\
&& F(i\omega)  \equiv  -\sum_{\mathbf{q}\nu}|V_\mathbf{q}|^2\int
d\tau e^{i\omega\tau}\nonumber \\
&& \langle
T_\tau[a_\mathbf{q}^\dag(\tau)+a_\mathbf{q}(\tau)]d_\nu(\tau)[a_\mathbf{q}^\dag(0)+a_\mathbf{q}(0)]d_\nu^\dag(0)\rangle.\label{eq:corr}
\end{eqnarray}
By assuming that the locations of the Kondo atoms $\mathbf{r}_i$ are
completely random, we can reduce the double momentum sum into a
single sum. The result for the differential conductance is
proportional to
\begin{equation}
\sum_{\mathbf{k}\sigma\mathbf{k'}\sigma'}\textrm{Im}[G_{\mathbf{k}\sigma\mathbf{k'}\sigma'}^r(\epsilon)]
\simeq-\pi\rho_0\left(1+\frac{n}{\rho_0D}\frac{\rho_0}{\pi}\textrm{Im}[F(\epsilon)]\right),\label{eq:sumG}
\end{equation}
where $D$ and $-D$ are set to be the cut-off energies above and
below the Fermi level. On the right hand side of
Eq.~(\ref{eq:sumG}), since $\rho_0$ gives the information on the
number of conducting channels of the bare nanowire, the additional
term in the parenthesis is the change caused by the Kondo atoms. The
part of $\rho_0\textrm{Im}[F(\epsilon)]$ is the effect contributed
by one single atom. The factor $n/\rho_0D$ is an interference
parameter due to the random locations of the Kondo atoms. This
factor is basically determined by the surface-to-volume ratio since
$n$ is the number of atoms on the surface and $\rho_0D$ gives the
total number of states on the wire. The surface-to-volume ratio here
also contributes to the average coupling constants $V$ and $V_0$
because the surface states couple more strongly to the atoms on the
surface.

\begin{figure}
\centering
\includegraphics[scale=0.4]{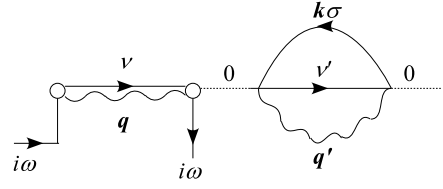}
\caption{Lowest non-trivial order Feynman diagram for $F(i\omega)$ in the $1/N$ expansion plotted with the same rules used in Ref.(\cite{Hewson}). The self energy correction for the left part of the diagram where there is one localized electron and one more or one less surface plasmon is neglected because of being at least in the order of $(1/N)^1$.}\label{fig:process}
\end{figure}
To evaluate the correlation function in Eq.~(\ref{eq:corr}), we
adapt the $1/N$ expansion with the approximation of
$U\rightarrow\infty$\cite{Hewson} to solve the photon-assisted
Anderson model. To demonstrate the feasibility, it is sufficient to
stop at the second order perturbation theory and keep only $(1/N)^0$
terms. The lowest non-trivial order Feynman diagram plotted in
Fig.~\ref{fig:process} and the calculation of Eq.~(\ref{eq:corr})
follows the procedures in Ref.\cite{Hewson}.

\begin{figure}
\centering
\includegraphics[scale=0.3]{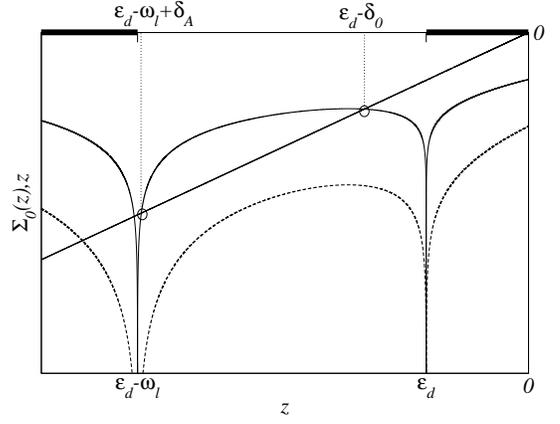}
\caption{Schematic graphical solution for the poles with the self energy of Eq.~(\ref{eq:SE}), $z=\Sigma_0(z)$. The solid curve and the dashed curve are the $\Sigma_0(z)$ for two different sets of parameters. For the solid curve, in addition to the original pole at $z=\epsilon_d-\delta_0$, another pole at $z=\epsilon_d-\omega_l+\delta_A$ appears corresponding to the process of electron hopping while absorbing a surface plasmon. The branch cuts, depicted as bold lines on the top, are chosen to be at $z<\epsilon_d-\omega_l$ and $\epsilon_d<z<\epsilon_d+\omega_l$, so that when the laser intensity $\alpha_l^2\rightarrow 0$ the system goes back to the original Anderson model. For the dashed curve, there is no pole lying outside the branch cuts, hence no Kondo resonance.}\label{fig:SE}
\end{figure}

Since $k_BT\ll\hbar\omega_\textrm{SPR}$, only the surface plasmon
states at the incident laser frequency are populated, and like its
laser source, is in a coherent state
$|\alpha_\mathbf{q}\rangle\delta_{\omega_\mathbf{q},\omega_l}$. Also
it is natural to assume that the laser frequency
$\omega_l=\omega_\textrm{SPR}$ and the surface plasmon states have
no degeneracy at the laser frequency, so that
$|\alpha_\mathbf{q}\rangle\delta_{\omega_\mathbf{q},\omega_l}=|\alpha_l\rangle$.
The self energy correction at zero temperature to the right part of
the diagram in Fig.~\ref{fig:process} is
\begin{eqnarray}
&& \Sigma_0(z)  = \frac{N_d\Delta_0}{\pi}\ln\left|\frac{z-\epsilon_d}{z-\epsilon_d-D}\right|+\frac{\alpha_l^2N_d\Delta}{\pi}\nonumber \\
&& \cdot
\left[\ln\left|\frac{z-\epsilon_d-\omega_l}{z-\epsilon_d-D-\omega_l}\right|
+\ln\left|\frac{z-\epsilon_d+\omega_l}{z-\epsilon_d-D+\omega_l}\right|\right]+ \frac{N_d\Delta}{\pi} \nonumber\\
&&  \cdot
\sum_\mathbf{q}\frac{(\kappa/2)^2}{(\omega_\mathbf{q}-\omega_\mathrm{SPR})^2+(\kappa/2)^2}\ln\left|\frac{z-\epsilon_d-\omega_\mathbf{q}}{z-\epsilon_d-D-\omega_\mathbf{q}}\right|,\label{eq:SE}
\end{eqnarray}
where $\Delta_0\equiv\pi\rho_0V_0^2$, $\Delta\equiv\pi\rho_0V^2$.
The first term on the right-hand side of Eq.~(\ref{eq:SE}) is the
self energy as in the original Anderson Model. The second term
corresponds to the electron hopping with stimulated surface plasmon
emission and absorption. The last term is the electron hopping while
spontaneously emitting a surface plasmon. The difference between the
result of the original Anderson model and Eq.~(\ref{eq:SE}) is that
there are two relevant branch cuts here. The branch cuts and the
graphical solution for finding the poles are plotted in
Fig.~\ref{fig:SE}. Assuming $D\gg\epsilon_d,\omega_l$, there are two
poles corresponding to the processes of electron tunneling with
absorbing a surface plasmon and without a surface plasmon. The case
of the dashed curve in Fig.~\ref{fig:SE} happens when the amplitude
of the incident laser is too large. Under this circumstances, the
Kondo effect is destroyed.

In the regime where the Kondo effect exists, the corresponding spectral function can be obtained by analytic continuation. The result for $T=0$ with the assumption of $\omega_l\gg \alpha_l^2N_d\Delta,N_d\Delta_0$ is
\begin{eqnarray}
&& \rho_0\textrm{Im}[F(\epsilon)]  \simeq
  -N_d\left[\Delta_0\left|1+\frac{N_d\Delta_0}{\pi\delta_0}\right|^{-1}\delta(\epsilon-\delta_0) \right. \nonumber\\
&& +\alpha_l^2\Delta\left|1-\frac{\alpha_l^2N_d\Delta}{\pi\delta_A}\right|^{-1}\delta(\epsilon+\delta_A)\nonumber\\
&&
+\alpha_l^2\Delta\left|1+\frac{N_d\Delta_0}{\pi\delta_A}\right|^{-1}[\delta(\epsilon-\omega_l-\delta_0)+
\delta(\epsilon+\omega_l-\delta_0)] \nonumber\\
&& +\Delta_0\left|1-\frac{\alpha_l^2N_d\Delta}{\pi\delta_A}\right|^{-1}\delta(\epsilon-\omega_l+\delta_A)\nonumber\\
&& \left.
+\alpha_l^2\Delta\left|1-\frac{\alpha_l^2N_d\Delta}{\pi\delta_A}\right|^{-1}\delta(\epsilon-2\omega_l+\delta_A)
\right].\label{eq:F}
\end{eqnarray}
In the measurement of nanowire differential conductance, the Kondo
effect reveals itself as anti-resonances, which corresponds to the
case with Fano parameter $q\rightarrow 0$\cite{Fano61,Madhavan01}.
In the presence of the strong surface plasmon field, the Kondo peak
near the Fermi surface splits into two due to the new opening of the
surface plasmon assisted tunneling. The side bands near
$\pm\omega_l$ and $2\omega_l$ also appear. The sizes of these
anti-resonance peaks can be controlled by varying the incident laser
intensity\cite{Shahbazyan00}. The reduction of the conductance at
the antiresonances is due to the electrons at these energies have a
resonant binding with the individual Kondo atoms and are unable to
conduct electricity.

\begin{figure}
\centering
\includegraphics[scale=0.65]{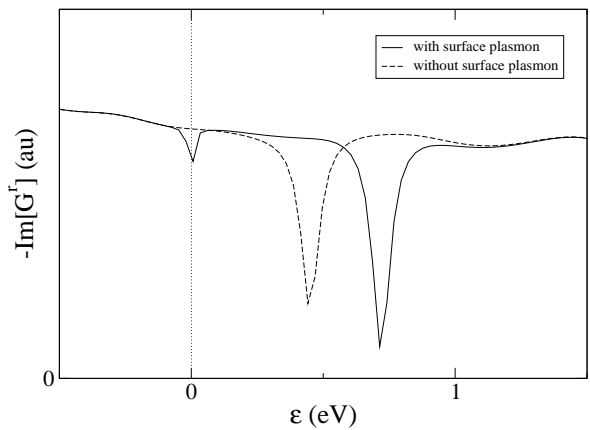}
\caption{The imaginary part of the retarded Green's function in
Eq.~(\ref{eq:J}) and Eq.~(\ref{eq:G}). The delta functions in
Eq.~(\ref{eq:F}) are artificially replaced by broadened Loretzian
functions of equal weight. The dashed curve shows the result of the
Co atoms adsorbed on gold nanowire, and the solid curve shows that
when the additional surface plasmon field is applied. It can be
clearly seen that the presence of the strong and high frequency (
$\omega_l>|\epsilon_d|$) EM field significantly shifts the original
Kondo peak to the right and an extra peak below the Fermi level
appears.}\label{fig:conductance}
\end{figure}

To estimate the feasibility, an {\em ab initio} calculation is
carried out to estimate the parameters. Our {\em ab initio}
calculation of the electronic structures of Co atoms interacting
with the Au surface and the $sp$-$d$ coupling strength caused by the
${\bf j}\cdot {\bf A}$ term is based on the full potential linear
augmented Slater-type orbital (LASTO) method developed by Davenport
and coworkers\cite{Daven1,Daven2,YC}. The LSDA+U scheme as described
in Ref. \cite{LDA+U} is adopted. It is believed that the LSDA+U
method is more reliable than the LSDA method in determining the
magnetization of strongly correlated systems. To study the $sp$-$d$
hybridization between the Co adatom and the Au surface, we adopt a
``supercell'' model which contains five layers of Au on an fcc
lattice (normal to the (111) direction) and three more vacuum
layers. One Co adatom is placed on the (111) surface of the Au slab
in each $\sqrt{3}\times\sqrt{3}$ surface unit cell (at equal
distances from the three Au atoms underneath). The surface-to-volume
ratio in this model is 1/5, so it corresponds to a nanowire of
radius $~ 10$ atomic layers or a diameter of $~10nm$ with 1/3
monolyer Co coverage on the surface. The on-site effective Coulomb
repulsion is taken to be $U\sim2.8\textrm{eV}$\cite{Ujsaghy00}. Our
LSDA+U calculation shows that the averaged occupancy of the Co $d$
orbitals is 7.92 (or a spin polarization of 2.08 $\mu_B$). With this
occupancy, we obtain $\epsilon_d\sim-1.18\textrm{eV}$,
$\Delta_0\sim0.4\textrm{eV}$ following the method used in Ref.
\cite{Ujsaghy00}. Assuming the incident laser of 525nm wavelength
has the focused intensity of $10^{11}\textrm{W/cm}^2$, whose
intensity of EM field is amplified by the surface plasmon resonance
by $10^3$, the resulting parameters
$\alpha_l^2\Delta\sim0.2\textrm{eV}$, $\delta_A\sim 2\textrm{meV}$,
and $\delta_0\sim0.72\textrm{eV}$. The $dI/dV$ curve is shown in
Fig.~\ref{fig:conductance}.

With the STM technology, the atoms on the wire can be arranged into
any desired pattern. Then this design can be used to investigate the
transport properties of 1D or 2D periodic Anderson model, since the
nanowire can be readily replaced by a thin film. The periodicity of
the Kondo atoms can be chosen at will, and the coupling parameters
of surface plasmon assisted tunneling can also be tuned by simply
varying the incident laser intensity. If realized, this could bring
the problem of traditionally microscopic strongly correlated
electron systems onto the surface of a mesoscopic scale.

In summary, we have proposed an experiment to measure the surface
plasmon assisted Kondo effect of magnetic atoms adsorbed on a
metallic nanowire and modeled the anticipated  $dI/dV$ curve via an
{\em ab initio} calculation. Due to the good surface-to-volume ratio
the electrons going through the nanowire are forced to scatter with
the Kondo atoms, which in our numerical calculation on a Co/Au case
gives a clearly observable effect. In addition, the introduction of
localized surface plasmon field on the metallic nanowire adds new
channels to the $sp$-$d$ hybridization so that the electrons now can
hop on and off the Kondo atom with absorbing or emitting real
photons, which results in multiple Kondo peaks.

The authors would like to thank Chii-Dong Chen, Yuh-Jen Cheng,
Kuo-Kan Liang, and Sheng-Hsien Lin for useful discussions. This work
was supported by Academia Sinica, Taiwan.
\bibliography{SEKR2}

\end{document}